\documentclass[11pt,preprint]{aastex}

\newcommand{\myemail}{hayato@crab.riken.jp}
\newcommand{\kte}{$kT_e$}
\newcommand{\net}{$n_et$}

\slugcomment{}

\shorttitle{Expansion Velocity of Ejecta in Tycho's SNR}
\shortauthors{Hayato et al.}

\begin{document}
\title{Expansion Velocity of Ejecta in Tycho's Supernova Remnant\\
Measured by Doppler Broadened X-ray Line Emission}

\author{Asami Hayato\altaffilmark{1,2},
Hiroya Yamaguchi\altaffilmark{1},
Toru Tamagawa\altaffilmark{1},
Satoru Katsuda\altaffilmark{2},
Una Hwang\altaffilmark{2},\\
John P. Hughes\altaffilmark{3},
Midori Ozawa\altaffilmark{4}, 
Aya Bamba\altaffilmark{5,6},
Kenzo Kinugasa\altaffilmark{7},
Yukikatsu Terada\altaffilmark{8},
Akihiro Furuzawa\altaffilmark{9}, \\
Hideyo Kunieda\altaffilmark{9}, and
Kazuo Makishima\altaffilmark{10,1}
}
\email{\myemail}

\altaffiltext{1}{RIKEN, 2-1 Hirosawa, Wako, Saitama 351-0198, Japan} 

\altaffiltext{2}{NASA Goddard Space Flight Center, Greenbelt, MD 20771, USA} 
  
\altaffiltext{3}{Department of Physics and Astronomy,  Rutgers University, 136 Frelinghuysen Road, Piscataway,  NJ 08854-8019, USA} 

\altaffiltext{4}{Department of Physics, Graduate School of Science, Kyoto University, Kita-Shirakawa, Sakyo-ku, Kyoto 606-8502, Japan}

\altaffiltext{5}{School of Cosmic Physics, Dublin Institute for Advanced Studies 31 Fitzwilliam Place, Dublin 2, Ireland}

\altaffiltext{6}{Institute of Space and Astronautical Science, Japan Aerospace Exploration Agency,   3-1-1 Yoshinodai, Tyuo-ku, Sagamihara, Kanagawa 252-5210, Japan}

\altaffiltext{7}{Gunma Astronomical Observatory, 6860-86, Nakayama, Takayama-mura, Agatsuma-gun, Gunma 377-0702, Japan}

\altaffiltext{8}{Department of Physics, Saitama University, Shimo-Okubo 255, Sakura, Saitama 338-8570, Japan} 

\altaffiltext{9}{Division of Particle and Astrophysical Science, Graduate School of Science, Nagoya University, Furo-cho, Nagoya 464-8602, Japan} 

\altaffiltext{10}{Department of Physics, The University of Tokyo, 7-3-1 Hongo, Bunkyo-ku, Tokyo 113-0033, Japan}

\begin{abstract}
We show that the expansion of ejecta in Tycho's supernova remnant
(SNR) is consistent with a spherically symmetric shell, based on
{\it Suzaku} measurements of the Doppler broadened X-ray emission lines.
All the strong K$\alpha$ line emission show broader widths at the center 
than at the rim,  while the centroid energies are constant across the remnant 
(except for Ca).
This is the pattern expected for Doppler broadening due to expansion of the 
SNR ejecta in a spherical shell.  To determine the expansion velocities of
the ejecta, we applied a model for each emission line feature having
two Gaussian components separately representing red- and blue-shifted
gas, and inferred the Doppler velocity difference between these two
components directly from the fitted centroid energy difference. Taking
into account the effect of projecting a three-dimensional shell to the
plane of the detector, we derived average spherical expansion
velocities independently for the K$\alpha$ emission of Si, S, Ar, and
Fe, and K$\beta$ of Si.  We found that the expansion velocities of Si,
S, and Ar ejecta of $4700\pm100$ km s$^{-1}$ are distinctly higher
than that obtained from Fe K$\alpha$ emission, $4000\pm300$ km
s$^{-1}$, which is consistent with segregation of the Fe in the inner
ejecta. Combining the observed ejecta velocities with the ejecta
proper-motion measurements by {\it Chandra}, we derived a distance to
the Tycho's SNR of $4\pm1$ kpc.

\end{abstract}


\section{Introduction}

Type Ia supernovae (SNe), the thermonuclear explosions of accreting
C$+$O white dwarfs, are important for their role as standard candles
to measure the distance to their host galaxies, and also as the main
sources for the production of the Fe-group elements. However, our
understanding of the physical mechanism of the explosion, as well as
the details of the progenitor systems, is incomplete. The kinetic
energy distribution of the ejecta holds important information about
the propagation of the supernova burning front and the resulting
nucleosynthesis.  In sufficiently young supernova remnants (SNRs),
this information is retained to some degree and can be probed by
studying the X-ray emission of the shocked ejecta.

Tycho's SNR is the remnant of the Galactic SN recorded by Tycho Brahe
in 1572, and is securely classified as a Type Ia SNR based on the
observed light curve \citep{baade45,ruiz04a} and the light-echo
spectrum \citep{krause08}. As the prototypical Ia SNR, Tycho's SNR has
been targeted for study of the explosion mechanism.  \citet{bad06} made
a comparison between spatially integrated X-ray spectra of Tycho's SNR
taken by {\it XMM-Newton} and {\it Chandra} and synthetic X-ray
spectra based on several different Type Ia explosion models. They 
concluded that the observed spectra are well reproduced by a one-
dimensional delayed detonation model with some degree of chemical 
stratification, with Fe-peak elements interior to intermediate-mass elements 
(IMEs, i.e., Si, S, Ar, Ca).

The narrow band images of Tycho's SNR obtained by {\it ASCA} show
that the Fe-K$\alpha$ emission is actually located interior to
the emission lines of IMEs and Fe-L emission \citep{hwang97}.
Furthermore, \citet{hwang98} found that the Fe-K emitting ejecta
have a higher electron temperature and a lower ionization age
than the IME ejecta. These results imply that the Fe-K emitting
ejecta are situated more inside of the remnant and has been
heated by the reverse shock more recently than the other ejecta.
Observations with {\it Chandra} \citep{warren05} and {\it
XMM-Newton} \citep{dec01} confirm the spectral segregation
between the IME and Fe.  The same situation has been reported for
other Ia SNRs such as SN 1006 \citep{yamaguchi08} and LMC SNRs
0509-67.5 and 0519-69 \citep{kosenko08,kosenko10}.

Thanks to the the spectral capability of {\it Suzaku}, \citet{furu09}
discovered a significant broadening of Fe K$\alpha$ line at the center
of the Tycho remnant. They concluded that the shell of Fe K$\alpha$
emitting ejecta is expanding with a line-of-sight velocity of
2800$-$3350 km s$^{-1}$. They also inferred line broadening for
He-like Si and S K$\alpha$~emission, but did not compute quantitative
velocities.  In this paper, we investigate the three-dimensional
structure of the ejecta in Tycho's SNR and determine the expansion
velocities.

\section{Observations and Data Reduction}

{\it Suzaku} observations of Tycho's SNR and an off-source background
were carried out on 2006 June 26-29 and 29-30, respectively, as a part
of the Scientific Working Group observing time. The pointing position
for the SNR was (RA, Dec) = (00$^{\rm h}25^{\rm m}20^{\rm s}$,
64$^{\circ}08\arcmin18\arcsec$); that for the background position was
(00$^{\rm h}36^{\rm m}54^{\rm s}$, 64$^{\circ}17\arcmin42\arcsec$),
1\fdg27 offset from the SNR along the Galactic plane.

{\it Suzaku} carries two active instruments: four X-ray Imaging
Spectrometers (XIS: Koyama et al. 2007) placed at the focal planes of
four X-ray Telescopes (XRT: Serlemitsos et al. 2007) and a non-imaging
Hard X-ray Detector (HXD: Takahashi et al. 2007).
Each XIS has a 17.$^{\prime}$8$\times$17.$^{\prime}$8 field
of view with a half-power diameter for the XRT of $\sim$
2$^{\prime}$. One of the four XIS sensors (XIS 1) is a
back-illuminated CCD with high sensitivity at $<$ 1 keV, while the
others (XIS 0, 2, and 3) are front-illuminated (FI) CCDs with high
efficiency and low background at $>$ 5 keV. All of the XIS sensors
were operated in the normal full-frame clocking mode without any
spaced-raw charge injection (SCI: Uchiyama et al. 2009) for both the
SNR and background observations\footnote{Tycho's SNR was observed
  again by {\it Suzaku} on 2008 August 4-8 and 11-12 as one of the
  large proposal programs of the third announcement of opportunity
  observing cycle, this time with SCI on. However, we report only the
  analysis of the 2006 data in this paper because the current
  calibration of the response for the SCI-off data is better than for
  the SCI-on data \citep{ozawa09}.}.

The HXD data sanalysis were already reported by \citet{tama09}, so we
present here the detailed analysis of the XIS data. We focus on only
the FI sensors (XIS 0, 2, 3), because these have better calibration of
the energy gain near the Si-edge.  We reprocessed the revision
2.0.6.13 data products using the {\tt xispi} software ({\tt HEASOFT}
version 6.5) with the version 20080825 {\tt makepi} file.
After the reprocessing, the data were cleaned with standard
screening criteria for the cleaned event data of the XIS. The
effective exposures of the SNR and background were 101 ks and 51
ks, respectively.

For the spectral fitting, we used response matrices created with
{\tt xisrmfgen} software version 2007-05-14 using the version
20080311 {\tt rmfparam} file. We estimated the accuracy of the
XIS energy scale for our own data and response functions using
the line centroid energies of the $^{55}$Fe calibration sources
at two of the four corners of each XIS chip. We fitted Mn
K$\alpha$~lines from $^{55}$Fe with Gaussian models and found
that the line centroids of XIS 0, 2, and 3 were all 0.1$-$0.14\%
higher than the expected energy of 5.985 keV.  This is somewhat
better than the systematic uncertainty of $\pm$0.2\% reported by
\citet{ozawa09}. We also checked the width of the Mn K$\alpha$ line
is fully consistent with the expected instrumental broadening,
even with all three FI sensors merged. The response files were generated 
by the {\tt xissimarfgen} software, assuming a homogeneous disk-like 
radiation source with a radius of 4$^{\prime}$ corresponding to the 
azimuthally averaged radius of Tycho's SNR measured from the {\it 
Chandra} image.

\section{Overall Features}
\label{sec:3}

\begin{figure}
\epsscale{.70}
\plotone{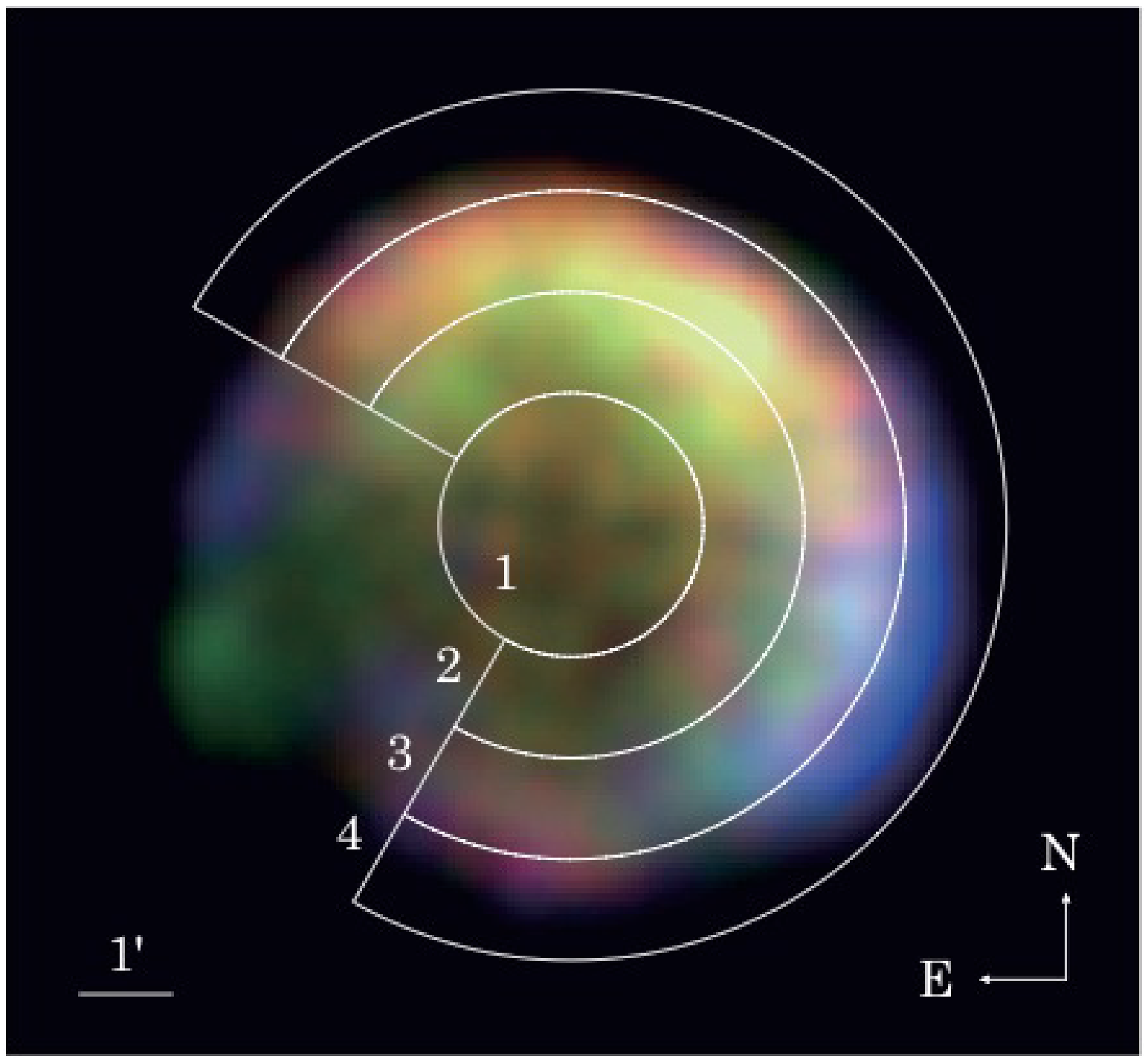}
\caption{Three color XIS FI (XIS 0+2+3) image of Tycho's SNR. Red, green, and blue correspond to the energy bands of Si He$\alpha$~(1.7$-$2 keV), the Fe K$\alpha$~(6.2$-$6.7 keV), and the hard band energy band (7$-$13 keV), respectively. The white circle and quarter sectors are the regions where we extract spectra.\label{fig:img}}
\end{figure}

A three-color XIS FI image of the SNR is shown in Figure \ref{fig:img}. Red, 
green, and blue colors correspond to the narrow energy bands of He-like Si 
K$\alpha$~(1.7$-$2 keV), Fe K$\alpha$~(6.2$-$6.7 keV), and the hard 
continuum band (7$-$13 keV). The images of the Si and Fe K$\alpha$ 
emission, which represent the ejecta distribution, are brighter in north and 
fainter in south, while the hard band image is brightest in the southwest. 
These agree with the trends seen in more detailed {\it Chandra} and {\it
XMM-Newton} images \citep{hwang02,warren05,bad06}.

\begin{figure}
\plotone{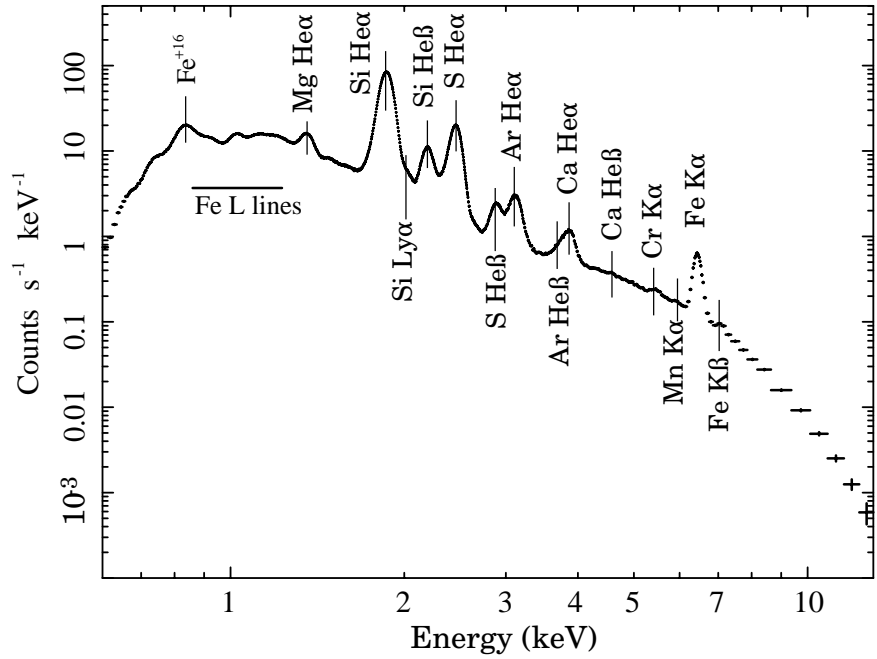}
\caption{Background-subtracted XIS FI (XIS 0+2+3) spectrum of the entire SNR. The important lines are marked. He$\alpha$, He$\beta$, and Ly$\alpha$~indicate emission lines of $n=2\rightarrow n=1$ transition in He-like ion, $1s3p\rightarrow 1s^{2}$ transition in He-like ion, and $n=2\rightarrow n=1$ transition in H-like ion. \label{fig:specall}}
\end{figure}

The background-subtracted XIS spectrum of the entire SNR is shown in
Figure \ref{fig:specall}. The source spectrum was taken from a circle
with a radius of 4\farcm65, with data from all three FI sensors merged
to improve photon statistics. The background spectrum was extracted
from a 7\arcmin-radius circular region on the off-source observation. In the 
spectrum, we can identify prominent K-shell emission line features of the 
He- and also some H-like ions of Mg, Si, S, Ar, and Ca, as well as emission 
from less ionized Cr, Mn and Fe
\citep{tama09}.

\section{Spatially-Resolved Spectra}
\label{sec:singlefit}

Figure 4 in \citet{furu09} clearly shows that the K$\alpha$ line blends
of Si, S, and Fe are broadened at the center of the remnant, relative
to the rim. However, their quantitative study was limited to the Fe
K$\alpha$ emission. Therefore, we extend their analysis to the lines
of IMEs. We divided Tycho's SNR into 4 radial regions as shown in
Figure \ref{fig:img}, with a circle at the center and three-quarter
circular rings numbered from 1 (inner) to 4 (outer). The radius of the
inner region 1 was 1\farcm41, and the thickness of the surrounding
outer regions was 1\farcm08. We excluded the southeast quadrant
(60\arcdeg$-$150\arcdeg~where the angles increase counterclockwise
from north) because of the presence of irregular ejecta clumps
\citep{vancura95,dec01}. We adopted the same background spectrum 
as used by \S\ref{sec:3}. Since the spectrum at energies below 1.7 keV is
dominated by the Fe L- shell emission, 
where the atomic physics is complicated, we focused only on the
energy band above 1.7 keV in the following sections. For simplicity,
we divided the spectrum into two energy bands: the 1.7$-$5 keV 
band for the lines of IMEs (\S\ref{sec:low}), and the 5$-$8 keV band
 for the Fe-K lines (\S\ref{sec:high}).

\subsection{1.7--5 keV Spectra}
\label{sec:low}

To examine the radial changes of emission features, we fitted spectra
taken from each region with a phenomenological model featuring an
absorbed power-law for the continuum emission plus twenty Gaussian
components for the line emission. The column density for the
foreground interstellar gas was set to be $N_{\rm H}=7\times10^{21}$
cm$^{-2}$ following \citet{cass07}. The Gaussians represented
transitions for the elements Si, S, Ar, and Ca for He-like and H-like
ions: He$\alpha$~($n=2\rightarrow n=1$), He$\beta$~($1s3p\rightarrow
1s^{2}$), He$\gamma$~($1s4p\rightarrow 1s^{2}$) in the He-like ions,
and Ly$\alpha$~($2p\rightarrow 1s$), Ly$\beta$~($3p\rightarrow
1s$) in the H-like ions. We modeled the He$\alpha$ transitions by a
single Gaussian, which actually contains a blend of the forbidden,
inter-combination, and resonance lines. For example, in Si He$\alpha$
blend, the rest energies of the constituent lines are 1839.4 eV (forbidden), 1853.7 eV
(inter-combination), and 1864.9 eV (resonance), with the blending 
giving an extra effective width of about 15 eV \citep{hwang97}. 
We also represented each blend of He$\beta$ and its surrounding
satellite lines by a single Gaussian model. We note that the
effective width for Si He$\beta$ is comparable to that of He$\alpha$,
based on the non-equilibrium ionization (NEI) model version 1.1 in
XSPEC.
The continuum spectra must be a combination of thermal and non-thermal 
emission, but these contributions are difficult to estimate separately (e.g., 
Tamagawa et al. 2009). We thus checked that the subsequent results for the 
Gaussian components do not change significantly even if we use a 
bremsstrahlung model for the continuum instead of a power-law model.

The line centroid, width, and intensity of the prominent He$\alpha$
blends of Si, S, Ar, and Ca, as well as the He$\beta$ blends of Si and
S, were fitted freely. The other emission lines are too weak to
constrain the Gaussian parameters. Thus, we fixed the energy
difference between the prominent line features above and weaker line
features of the same element; for example, the energy difference
between S He$\alpha$ and S Ly$\alpha$ was fixed to 178 eV. Here we
assumed electron temperature \kte~of $\sim$ 1 keV and ionization
timescale \net~of $\sim10^{11}$ cm$^{-3}$ s, typical in Tycho's SNR
\citep{hwang98}, because the centroid energy of the He$\alpha$ 
blend depends on \kte~and \net; and used the 
Astrophysical Plasma Emission Database (APED: Smith et al. 2001).
The line widths were linked to each other for lines of the same element
(e.g., S He$\alpha$, S Ly$\alpha$, S He$\beta$, S He$\gamma$, 
and S Ly$\beta$). The intensities of the prominent He-like emission above 
and the relatively strong Ly$\alpha$ lines of Si and S, were freely fitted. For 
lines other than those, we fixed the intensity ratios: we took
He$\gamma$/He$\beta$=0.3 and Ly$\beta$/Ly$\alpha$=0.1 for the plasma
in Tycho's SNR (\kte=1 keV and \net=10$^{11}$ cm$^{-3}$ s, as noted
above). These intensity ratios do not vary much with \net~and vary
only 10\% over a decade in \kte; hence the uncertainties here are
negligibly small. We also fixed the ratios of Ly$\alpha$/He$\alpha$
and He$\beta$/He$\alpha$~for Ar and Ca to be the same as these of S.

\begin{figure}
\epsscale{.80}
\plotone{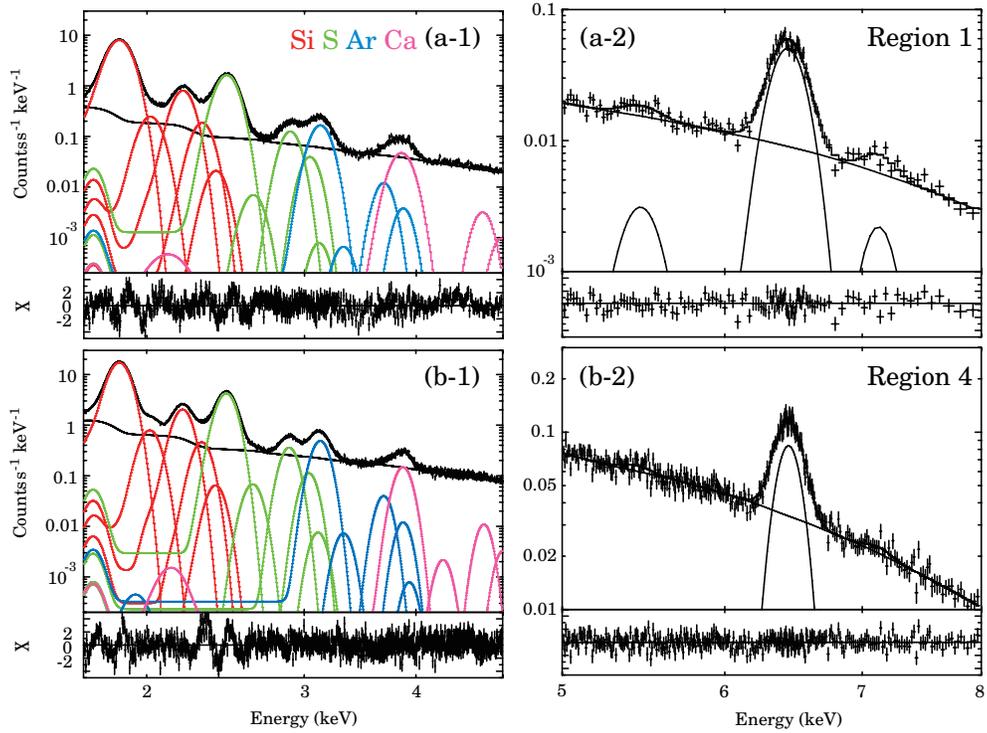}
\caption{Background-subtracted 1.7$-$5 keV (left) and 5$-$8 keV (right) band spectra extracted from region 1 (a) and 4 (b) indicated in Figure \ref{fig:img}. The best-fit models are shown in solid lines and residuals are given in the bottom panels. \label{fig:single}}
\end{figure}

With these models and assumptions, we fitted all the four spectra. The
reduced $\chi^2$ values and degrees of freedom (dof) in regions 1, 2, 3, 
and 4 were 1.37 (791), 1.42 (724), 1.74 (858), and 1.56 (846), respectively. 
The large reduced $\chi^2$ values are attributed to the high statistics of 
spectra and our approximations above for the fitting model. In addition, 
there might be some contaminating emission from the less ionized ions 
(e.g., Li-like). We thus conclude that the best-fit models do reproduce the 
spectra sufficiently well. Individual components of the best-fit models for 
regions 1 and 4 are shown in Figure~\ref{fig:single} (a-1) and (b-1), 
respectively. The best-fit parameters of all the fits are listed in 
Table~\ref{tab:single}.

\subsection{5--8 keV Spectra}
\label{sec:high}

We also fitted the $5-8$ keV band spectra with a power-law for the
continuum, plus three Gaussian components for Cr and Fe K$\alpha$, and
Fe K$\beta$ line blends. Since Cr K$\alpha$~and Fe K$\beta$ emission
are very weak, these widths were tied to that of Fe K$\alpha$. The other 
parameters were allowed to vary freely. The reduced $\chi^2$s and dof for 
regions 1, 2, 3 and 4 were 1.25 (115), 1.22 (214), 1.07 (335), and 0.86 
(298), respectively.  All the fits for the four spectra from regions 1$-$4 were 
acceptable. The 5$-$8 keV spectra from regions 1 and 4 with the best-fit 
models are shown in Figure \ref{fig:single} (a-2) and (b-2), respectively. The 
best-fit parameters are listed in Table \ref{tab:single}.

\section{Radial Line Profiles}
\label{sec:rp}

\begin{figure}
\epsscale{1.0}
\plotone{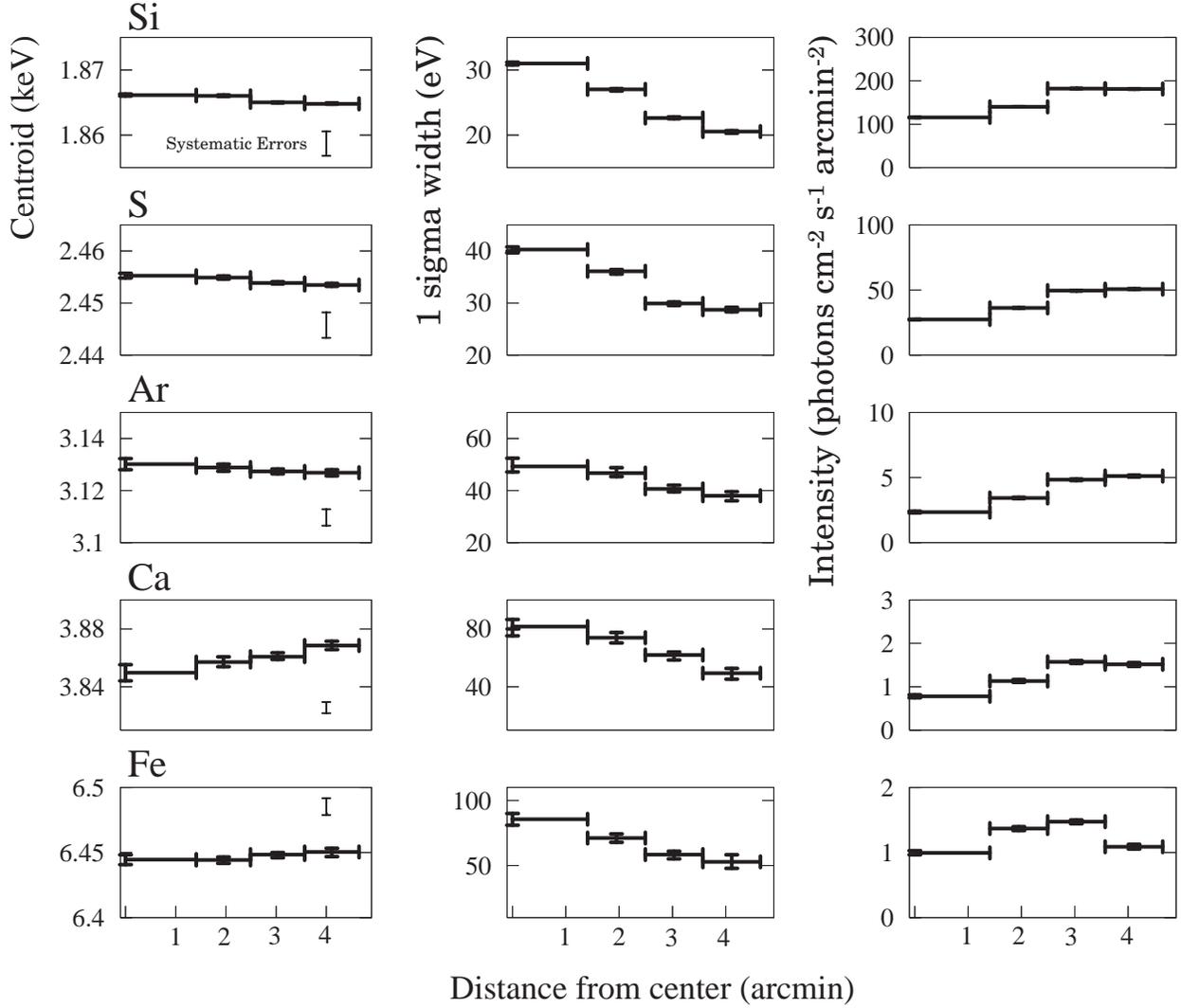}
\caption{Radial dependence of the best-fit Gaussian parameters of He$\alpha$~lines of Si, S, Ar, and Ca, and Fe K$\alpha$ blend. The centroid energies (left), the 1 $\sigma$ widths (middle), and the   intensities (right).  The systematic errors from the intrachip gain variation at 90\% confidence level are represented at the right corner in each left box. The statistical errors also represent 90\% confidence level. \label{fig:rp}} 
\end{figure}

Figure \ref{fig:rp} shows the centroid energies, widths, and intensities of the 
K$\alpha$ line blends of Si, S, Ar, Ca and Fe, obtained in the previous 
section.  We found that the centroid energies of all the elements except for 
Ca are constant with respect to radius within the systematic uncertainties of 
$\pm0.2$\%, indicated in each left box. This systematic error was estimated 
by \citet{ota07} using the {\it Suzaku} observations of extended sources with 
Fe K$\alpha$ emission (Cygnus Loop and Sagittarius C), and represents 
the gain variation on the same CCD chip at the 90\% confidence level. In 
contrast to the constancy of the centroid energies, the widths of all the 
emission features decrease significantly from the center to the rim.  The 
difference between the Si and Fe widths of regions 1 and 4 is $10.5\pm0.3$ 
eV and $33\pm6$ eV, respectively.

The radial intensity profile of IMEs emission are similar to each
other with a peak radius of $3\arcmin-4\arcmin$, while Fe-K$\alpha$
emission has a radius of $\sim3$\arcmin~which is somewhat smaller than
that of the IMEs. These trends agree with the previous observations
(e.g., Hwang \& Gotthelf 1997).

There is some indication of spectral variations within our source
regions in Figure \ref{fig:img}, so we investigated the azimuthal variations of
the line centroids and widths by dividing regions 2, 3, and 4 evenly into
three azimuthal sectors. We applied the same fitting procedure to the
spectrum accumulated from each divided region. There is some variation
amongst the azimuthally divided sectors, with the azimuthal variations
of line centroids being $\pm3$ eV for Si, $\pm5$ eV for S, $\pm8$ eV
for Ar, $\pm20$ eV for Ca, and $\pm10$ eV for Fe.  These values are
much smaller than the corresponding line widths; however, in all
the radial sectors, the widths show the same trends of being wider in
the inside compared to the rim, and the centroids being more constant.
The qualitative trends appear to be the same as for the azimuthally
averaged spectral regions, therefore we proceed with our
azimuthally averaged analysis.

In the following subsections, we consider two simple interpretations
of the radial properties of centroid energies and widths as discussed
by \citet{furu09}: (a) plasma characterized by multiple ionization
ages, and (b) Doppler shifts resulting from expansion in a shell.

\subsection{Plasma with Multiple Ionization Ages}

It is expected that the plasma in the SNR will have multiple
ionization ages. Since the reverse shock propagates from the outside
of the remnant inward, the shocked ejecta will have different \net~between
the outside (near the contact discontinuity) and the inside (where the
reverse shock has just passed through), with the more recently shocked 
ejecta generally having a lower \net~and a different \kte~from the outer 
ejecta, with details depending on the ejecta density profile
\citep{dwarkadas98}.

Plasma with a range of \net~values easily makes the observed line blend
broaden at the center of the remnant, since line centroids strongly depend
on $n_et$. For example, based on the NEI model of version 1.1, the
centroid of the K$\alpha$ line blend of S at \kte=1 keV varies from 2.41 keV
to 2.45 keV depending on the \net~values of $10^{10}-10^{11}$ cm$^{-3}$ s,
which may result in a line broadening of $\sim$40\,eV. However, our results
show that the centroid energies are constant across the remnant within 
0.2\%. This implies that the plasma emitting each line or blend has the same
average \net~all over the SNR.  Therefore, a multiple \net~plasma does not 
provide a fully satisfying explanation of our observation.
 
\subsection{Doppler Shift by the Shell Expansion}

Another case we consider is an expanding shell of ejecta.
The emission lines from retreating and approaching gas are red- and blue-
shifted, respectively.  Thus, the line emission at the projected center of the 
remnant is expected to be broadened. At the rim, on the other hand, a 
narrow line would be observed, since there is only a small line-of-sight 
component of the velocity.  In addition, if the expansion is spherically 
symmetric (and the plasma is reasonably uniform), the centroid energies will 
be constant with radius. We found that the observed line profiles are well 
reproduced by the Doppler broadening of the spherically symmetric shell 
expansion. 

\subsection{Properties of Ca He-like K$\alpha$~line}
\label{sec:ca}

As shown in Figure \ref{fig:rp}, the Ca He$\alpha$ blend is the only
feature where the centroid energy is seen to gradually increases from
the center to the rim. Possibly, this suggests that the Ca ejecta might be in 
multiple ionization ages unlike the other ejecta. However, the Ca 
He$\alpha$ blend is heavily contaminated by Ar He$\beta$, and thus the 
inferred properties of the Ca emission depends rather strongly on the 
assumed intensity ratio, He$\beta$/He$\alpha$ of Ar.  A future 
observation with higher energy resolution is required to reliably investigate 
the accurate properties of Ca He$\alpha$ blend.

\section{Velocities of Ejecta}

In this section, we quantitatively derive the expansion velocity with
the K$\alpha$ emission of Si, S, Ar, and Fe, and K$\beta$ of Si,
assuming that the ejecta are expanding in a spherical shell.

\subsection{Method of Velocity Determination}

In a spherically symmetric shell expansion, a broadened line should have
both red- and blue-shifted components. If we apply such a two-Gaussian
model (with red- and blue-shifted components) to a single broadened
line, the line energy shift (i.e., relative to the rest frame) gives
the Doppler velocity as
\begin{equation}
\label{eq:doppler}
\frac {\left| E_{{\rm obs},i}-E_0\right|}{E_0}=\frac{v_{\perp i}}{c}~,
\end{equation}
where $E_{{\rm obs},i}$ and $v_{\perp i}$ are the observed
centroid energy of the  red- or blue-shifted line and the line-of-sight velocity
in each region $i~(i=1-4)$, respectively, and $E_0$ is the line energy at the 
rest frame. We introduce here the energy shift in each region $i, \delta
E_{i}=\left|E_{{\rm obs},i}-E_0\right|$. Assuming front-back velocity
symmetry, the parameter we measure is $2\times\delta E_{i}$,
corresponding to the energy difference between the red- and
blue-shifted lines.

Once we obtain $v_{\perp i}$, we can convert it to the expansion
velocity $v_{\rm exp}$ by considering two important effects: the
projection of the three-dimensional shell onto the plane of the
detector, and the limited spatial resolution of the {\it Suzaku} XRT.
In an appendix, we calculate the factors $C_i$ which represent what
percentage of the $v_{\rm exp}$ would be observed as $v_{\perp i}$
in each region $i$. In other words, $v_{\rm exp}$ can be expressed in 
terms of $C_i$ and the observable parameter $v_{\perp i}$ as
\begin{equation}
\label{eq:cf}
v_{\rm exp}=\frac{v_{\perp i}}{C_i}~.
\end{equation}
The calculated $C_i$ factors are summarized in Table \ref{tab:cf}.

\subsection{Spectral Fitting}
\label{sec:double_fit}

As a first step, we derived the expansion velocities by fitting the
spectra of regions 1 and 4 at the same time, because the difference 
between $v_{\perp 1}$ and $v_{\perp 4}$ is expected to be the largest.

To test for spherically symmetric expansion,
we therefore fitted the 1.7$-$5 keV spectra with
the model of an absorbed power-law for the continuum and Gaussian
components for emission lines. The column density for the foreground
interstellar gas was set as in \S\ref{sec:low}. We applied Gaussian
models including red- and blue-shifted components, for each line
feature.  We left the centroid energies of the both lines free, while
the widths and intensities were set to be equal to each other. Because
we fitted the spectra of regions 1 and 4 at the same time, we actually
applied two pairs of the red- and blue-shifted lines (four Gaussians
in total) for each line feature. According to Equation
\ref{eq:doppler}, the ratio of $\delta E_1/\delta E_4$ equals the
ratio of $v_{\perp 1}/v_{\perp 4}$. Since $v_{\rm exp}$ should be
the same for regions 1 and 4, Equation \ref{eq:cf}
gives $\delta E_1/\delta E_4 = C_1/C_4$. In our fitting, we therefore
fixed the ratio of $\delta E_1/\delta E_4=C_1/C_4$. As a result,
$v_{\rm exp},$ the width, and intensity were the only three parameters
that were allowed to vary freely for each set of four Gaussians.

We employed twenty sets of four Gaussians for the observed line
features: He$\alpha$, Ly$\alpha$, He$\beta$, He$\gamma$, Ly$\beta$ of
Si, S, Ar, and Ca. We derived the $v_{\rm exp}$ independently for the
prominent He$\alpha$ blends of Si, S, and Ar, and He$\beta$ of Si. The
energy separation between these and other line features were fixed as
the model in \S\ref{sec:low}; for example, the energy difference
between Ar He$\alpha$ and Ar He$\beta$ was fixed to 209 eV (see Table
\ref{tab:single}). The sole exception is S Ly$\alpha$, whose energy
was free in \S\ref{sec:low}; here, we fixed the energy shift from S
He$\beta$ to 181 eV to compensate for limited photon statistics. A
single line width was fitted for lines of the same element (e.g., for
S: He$\alpha$, Ly$\alpha$, He$\beta$, He$\gamma$, and Ly$\beta$).
The exception was Si He$\alpha$, whose width was fitted independently
from that of other Si lines, due to uncertainties in the response near
the Si-edge. 
As noted in \S\ref{sec:low}, the width of the Gaussian applied to the
He$\alpha$ blend includes a contribution from the blending of the 
constituent triplet lines ($\sim15$ eV). This broadening varies within a few 
eV depending on the plasma condition, but we checked that the 
measurement of the line centroid is little affected by the uncertainty of the 
width.
The intensity ratios of weak lines to prominent
lines (e.g., S Ly$\alpha$/He$\alpha$) were fixed by assuming a plasma
with \kte$\sim$ 1 keV and \net$\sim10^{11}$ cm $^{-3}$ s, as in
\S\ref{sec:low}.  The exception here is that S Ly$\alpha$/He$\alpha$
was fixed to the value obtained from the fits in \S\ref{sec:low}
(Table \ref{tab:single}).

\begin{figure}
\epsscale{.90}
\plotone{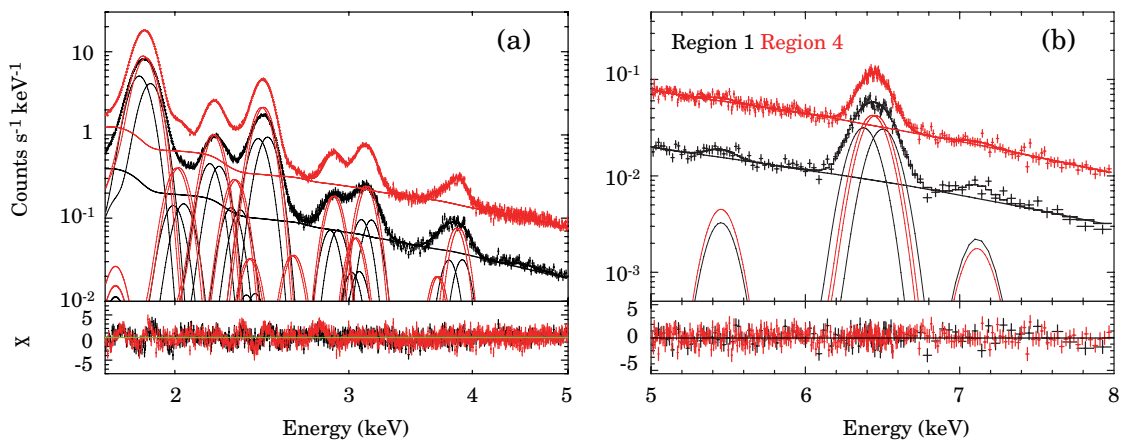}
\caption{(a) 1.7$-$5 keV and (b) 5$-$8 keV spectra of region 1 (black) and 4 (red) with the best-fit double Gaussian models. \label{fig:double14}}
\end{figure}

We also fitted the $5-8$ keV spectra of regions 1 and 4 using the same
procedure as above. The model consisted of a power-law continuum and
red- and blue-shifted Fe K$\alpha$ lines. We also included the Cr
K$\alpha$~and Fe K$\beta$ blends, but as single Gaussians due to
limited photon statistics.  We fixed the ratio of $\delta E_1/\delta
E_4$ to $C_1/C_4$ following Table \ref{tab:cf}. The widths and
intensities of the red- and blue-shifted lines in regions 1 and 4 were
all linked to each other.  Figure \ref{fig:double14} shows the
individual components of the best-fit models. 
The reduced $\chi^2$s and dof were 1.47 (1430) for 1.7$-$5 keV and 0.96 
(415) for 5$-$8 keV band spectra.

As a second step, we have also fitted all the spectra of regions 1, 2,
3, and 4 at the same time in exactly the same manner just described.
Here our aim was to verify that the model was valid for all four
regions, and also to improve the photon statistics.  The ratios of
energy difference among regions 1, 2, 3 and 4 were fixed as $\delta E_1
: \delta E_2 : \delta E_3 : \delta E_4 = C_1 : C_2 : C_3 : C_4$.  The
parameters obtained from the fits are consistent with those obtained for
regions 1 and 4, and within the systematical errors. The reduced
$\chi^2$s and dof were 1.55 (3028) and 1.06 (970) for 1.7$-$5 keV and
5$-$8 keV band spectra, respectively. The best fit parameters are
summarized in Table \ref{tab:double1234}.

\subsection{Expansion Velocities}

We translate $\delta E_{i}$ into $v_{\perp i}$ using Equation
\ref{eq:doppler}, and then convert to $v_{\rm exp}$ using Equation
\ref{eq:cf} and $C_i$ (Table \ref{tab:cf}). The derived velocities are
summarized in Table \ref{tab:double1234}.
We note that the $v_{\rm exp}$ of the Fe K$\alpha$ emission in 
Table \ref{tab:double1234} does not correspond to the line-of-sight velocity
measured in \citet{furu09}, because the $v_{\rm exp}$ represents the
expansion velocity in which the projection effect and PSF effect
have properly been taken into account. 
Meanwhile, the $v_{\perp 1}$ is the value should correspond to 
the line-of-sight velocity measured in \citet{furu09},
$3040^{+310}_{-240}$ km s$^{-1}$.
Our measurement of $v_{\perp 1}$ and the line-of-sight velocity
measured in \citet{furu09} agree well each other.
The $v_{\rm exp}$ derived from Si He$\alpha$ and
He$\beta$ are consistent with each other. Therefore, we conclude
that the systematic uncertainty of the Si-edge in the response is
small enough to determine the velocity.  The expansion velocity 
obtained from the He$\alpha$ blends of Si, S, and Ar are clearly higher
than that obtained from the Fe K$\alpha$ blends.

\section{Discussion}

Based on the analysis of spatially resolved spectra, we have demonstrated that
(1) the ejecta shell is expanding in a generally spherical and symmetric manner,
and (2) the expansion velocities of the IME ejecta are significantly higher than
that of the Fe K$\alpha$ emitting ejecta. These results then allow us to discuss
the distance to Tycho's SNR, the ejecta segregation, and the reverse shock velocity.

\subsection{Distance to Tycho's SNR}

Using {\it Chandra} high-resolution images of Tycho's SNR obtained in
2000, 2003, and 2007, \citet{katsuda09} measured the expansion rates of
both the forward shock and the reverse shocked ejecta. They found that
the mean proper-motion of the Si-rich layer is $\mu \sim
0\farcs25$~yr$^{-1}$. This is consistent with the result \citet{hughes00} 
had derived using {\it ROSAT} data. Combining with our expansion velocities of
$4700\pm100$ km~s$^{-1}$ (Table \ref{tab:double1234}), we obtained a
range of the distance to the SNR of $D = (4.0 \pm 1.0) (v/4700$ km
s$^{-1}$) ($\mu /0\farcs25$ yr$^{-1}$)$^{-1}$ kpc. The relatively
large uncertainty is mainly due to the azimuthal variation of the
proper motion of the ejecta. Our result is the first estimate of the
distance to Tycho derived solely from X-ray observations.

\begin{figure}
\epsscale{.70}
\plotone{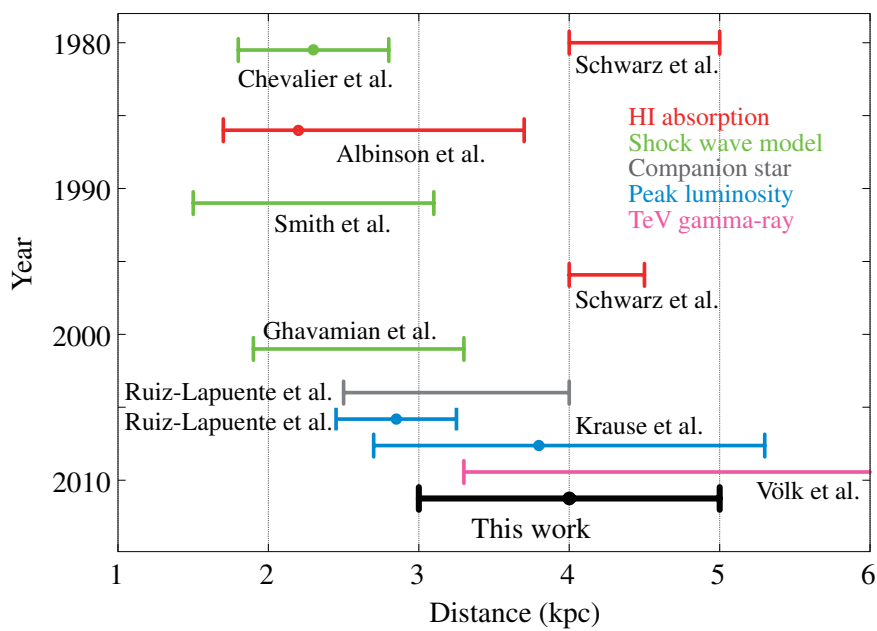}
\caption{Estimates of the distance to Tycho's SNR. \label{fig:dis}}
\end{figure}

The distances inferred by previous studies is shown in Figure
\ref{fig:dis}. The distance of around 2$-$3 kpc estimated by modeling
the observed H$\alpha$ line spectra (green in Figure \ref{fig:dis})
has been most widely cited thus far, but is model-dependent.
\citet{krause08} recently derived a larger distance of
3.8$^{+1.5}_{-1.1}$ kpc, based on the SN peak luminosity estimated by
the observed optical light-echo spectrum. Our estimate is consistent
with the result from \citet{krause08}.

\subsection{Ejecta Segregation}

The Fe-K radial profile peaks at $\sim180$\arcsec~which is a
distinctly smaller radius than the IME and Fe-L emission of
$\sim200$\arcsec~\citep{dec01,warren05}.  In addition, the Fe-K
emitting ejecta has about 100 times lower \net~than the IMEs and Fe-L
emitter (10$^{11}$ cm s$^{-1}$: Hwang et al. 1998). These results
suggest that the plasma emitting Fe-K has been heated by the reverse
shock more recently than the other materials. Since the reverse shock
propagates from the outside toward the center, the Fe-K and Si
emission are tracers of the interior and exterior material in the SNR,
respectively.

Our results show that the Si ejecta also have higher $v_{\rm exp}$
($4700\pm100$ km s$^{-1}$) compared to the Fe K$\alpha$ emitting
ejecta ($4000\pm300$ km s$^{-1}$). 
These velocity measurements add to
the morphological and spectral evidence that the Fe-K emitting ejecta are
segregated interior to the Si ejecta. 
The expansion rate ratio of Fe
K$\alpha$ emission to Si ejecta is derived to be
$(4000/180)/(4700/200)\sim0.93$.

\subsection{Reverse Shock Velocity}

The angular radius of the reverse shock of 183\arcsec~estimated by
\citet{warren05} corresponds to $r_{\rm rs}=3.5\pm0.9$ pc at a
distance of $4.0\pm1.0$ kpc. Given the age of 434 yr, the velocity of
unshocked ejecta $v_{\rm un, ej}$ is then estimated to be $8000\pm2000$
km s$^{-1}$ at a radius of 3.5 pc. The post-shock velocity $v_{\rm sh,
  ej} $ can be approximated as the obtained expansion velocity of Fe,
$4000\pm300$ km s$^{-1}$.
Under the assumption of strong shock, conservation laws for mass, momentum, and energy across
the shock front yield the following Rankine-Hugoniot relation,
\begin{equation}
\label{eq:rh}
v_{\rm un,ej}-v_{\rm rs}\,=\,
\left(\frac{\gamma +1}{\gamma -1}\right)
\left(v_{\rm sh,ej}-v_{\rm rs}\right),
\end{equation}
where $v_{\rm rs}$ and $\gamma$ are the reverse shock velocity in the
observer's frame and specific heat ratio, respectively. Since
\citet{warren05} argued that the reverse shock of Tycho's SNR is not
strongly accelerating cosmic-rays (unlike the forward shock), we here
take $\gamma=5/3$ as appropriate for a non-relativistic ideal
gas. Equation \ref{eq:rh} then gives $v_{\rm rs}=2700\pm800$ km
s$^{-1}$, and hence the upstream velocity in the shock-rest frame is
obtained to be $\bar{v}_{\rm rs}=v_{\rm un,ej}-v_{\rm rs}=5300\pm2100$
km s$^{-1}$.

This value allows us to estimate the explosion energy of Tycho's SN by
comparing with a self-similar model for the evolution of young SNRs
interacting with a uniform density ISM. In Figure 2d of
\citet{dwarkadas98}, $\bar{v}_{\rm rs}$ is expected to be
$3.4\times10^3~E^{1/2}_{51}(M_{\rm ej}/1.4M_{\sun})^{-1/2}$ km
s$^{-1}$ throughout most of the SNR's evolution, for either $r^{-7}$
power-law and exponential density distributions of the ejecta, where
$E_{51}$ is the explosion energy in units of $10^{51}$ ergs. Thus,
assuming an ejecta mass of 1.4$M_{\sun}$, the explosion energy of
Tycho's SN is derived to be $(2.5\pm2.0) \times 10^{51}$ ergs. Given
that the error is large, the derived energy range includes the
standard value for normal Type Ia SNe.

From the Rankine-Hugoniot relations, the post-shock temperature for
each particle species $a$ is given as
\begin{equation}
  \label{eq:temp}
  kT_a = \frac{2(\gamma - 1)}{(\gamma + 1)^2} m_a \bar{v}_{\rm rs}^2 
  = \frac{3}{16} m_a \bar{v}_{\rm rs}^2, 
\end{equation}
where $m_a$ is the particle mass. Given $\bar{v}_{\rm rs}=5300\pm2100$
km s$^{-1}$, the temperature of the shocked Fe is expected to be
$kT_{\rm Fe}=3\pm2$ MeV. Such
a high temperature would result in a line broadening of $\Delta
E=70\pm30$ eV due to the thermal Doppler effect. We measured the width
of the Fe-K blend of 55$\pm3$ eV (Table \ref{tab:double1234}), where an
intrinsic line broadening of $20-30$ eV is expected for the plasma in
the range of \kte$>1.6$ keV and $10<$ log \net~cm$^{-3}$ s $<10.7$
\citep{hwang98,furu09}. The additional width of $\sim$50 eV
which cannot be explained by the intrinsic line broadening
might be explained by the thermal Doppler broadening of Fe ions. Future
missions with high energy resolution, like $Astro$-$H,$ will help us
to accurately measure ion temperatures and study the heating process
at the reverse shock of young SNRs in detail.

\section{Summary}

We analyzed the X-ray spectra of Tycho's SNR using the {\it Suzaku}
100 ks observation. We summarize our results below:

\begin{enumerate}
\item We obtained the radial dependence of centroid energies, widths,
  and intensities for the K$\alpha$ emission of Si, S, Ar, Ca, and Fe
  for annular regions covering nearly three-fourths of the azimuth
  around the SNR. The centroid energies of all line blends except Ca
  are constant with radius, while the widths significantly become
  narrower from the center to the rim.  We found that the observed
  line properties are well explained by a spherically symmetric shell
  expansion of the ejecta.
\item We derived the expansion velocities of ejecta with Doppler
  broadened K$\alpha$ emission of Si, S, Ar and Fe, and K$\beta$ of Si
  independently. We fitted each broadened line feature with two
  Gaussians representing red- and blue-shifted gas, and obtained the
  expansion velocity from the centroid energy separation between two
  lines. We found that expansion velocities measured for K$\alpha$
  emission of Si, $4700\pm100$ km s$^{-1}$, are clearly higher than
  that measured for Fe K$\alpha$, $4000\pm300$ km s$^{-1}$.
\item Combining the obtained expansion velocity with the proper-motion
  measurement by {\it Chandra}, the distance to Tycho's SNR is
  estimated to be $4\pm1$ kpc.

\end{enumerate}

\appendix

To determine the true expansion velocity $v_{\rm exp}$, we have to
calculate the correction factors $C_i$ introduced in Equation
\ref{eq:cf}. $C_i$ describes the projection effect of the
three-dimensional shell onto the detector plane, including the photon
smearing effects caused by the limited spatial resolution of the {\it Suzaku}
XRT. In this appendix, we calculate the correction factors
$C_i$.

\begin{figure}
\epsscale{.50}
\plotone{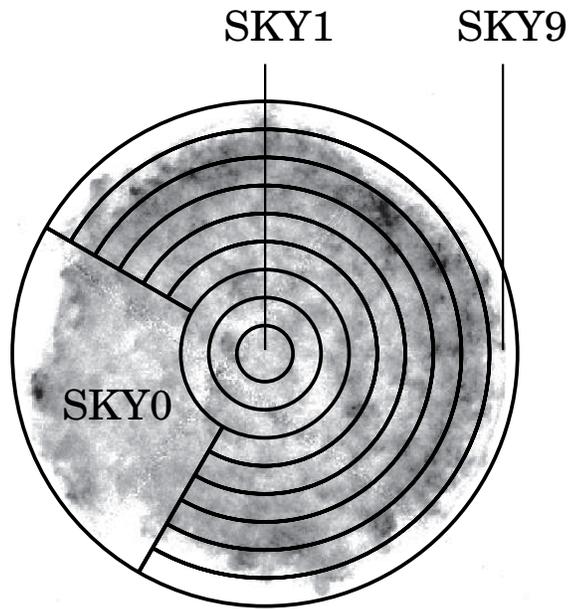}
\caption{SKY regions defined in the plane of the sky. The underlying image was taken by {\it Chandra}.\label{fig:regions}}
\end{figure}

First we define the regions that we use in the following
calculations. As shown in Figure \ref{fig:regions}, we divide the
entire SNR into ten ``SKY" regions. The region enclosed by the
innermost circle is named SKY 1, the next two annular regions are SKY
2 and 3, and the outer six annular regions, in which a quarter region is
excluded, are named SKY 4 (inner) to SKY 9 (outer).  The radii of the
SKY regions are 30\arcsec~for SKY 1, and are incremented by
30\arcsec~toward the outer edge of the SNR. The quarter region in the
southeast is named SKY 0. The regions $1-4$ are the detector regions
used in \S\ref{sec:singlefit} (Figure \ref{fig:img}). The regions 1, 2,
3, and 4 roughly correspond to the SKY $1-3$, $4-5$, $6-7$, and $8-9$,
respectively.

\section{Projection Effects}

Consider a thin and spherically symmetric shell with radius
$r_{\rm sh}$. The shell is expanding with velocity $v_{\rm exp}$. The polar 
angle $\theta$ is defined as the angle measured from the horizontal line which 
intersects the center of the spherical shell, and the azimuth angle $\phi$
represents an angle around the horizontal line. The sky region
$j~(j=0, 1, \centerdot\centerdot\centerdot, 9)$ is described by using
the range of $\theta$ ($\theta ^j_1$ and $\theta ^j_2$) and $\phi$.
The $\theta^j_1$ and $\theta^j_2$ angles are given as
\begin{eqnarray}
\theta^{0}_1 & = & \arcsin\left(r_3/r_{\rm sh}\right), \,\theta^{0}_2  =  \arcsin\left(r_9/r_{\rm sh}\right), \\
\theta^{j}_1 & = & \arcsin\left(r_{j-1}/r_{\rm sh}\right),\, \theta^{j}_2 =  \arcsin\left(r_j/r_{\rm sh}\right)
  (j=1, 2, \centerdot\centerdot\centerdot, 9),\\
\end{eqnarray}
where the radius $r_k$ is given as
\begin{equation}
r_k=k\cdot30\arcsec ~(k=0, 1, \centerdot\centerdot\centerdot, 9).
\end{equation}
We here consider only the half of the spherical shell ($0<\theta<90$ deg) in 
to evaluate $\delta E_{i}$. 

The projected velocities $v_j$ of the SKY region $j$ are found as the
mean line-of-sight velocity $v_{\perp}=v_{\rm exp}\cos{\theta}$. By integrating 
$v_{\perp}$ over the solid angle, the mean velocity $v_j$ is  described as 
\begin{equation}
\label{eq:vmean}
 v_j=v_{\rm exp}\cdot A_j\,,
\end{equation}
where
\begin{equation}
A_j=\frac{\displaystyle
  \int_{\theta^l_1}^{\theta^j_2}\cos\theta\,\sin\theta\,d\theta}{
  \displaystyle\int_{\theta^j_1}^{\theta^j_2}\sin\theta\,d\theta}~.
\end{equation}
Since $\theta^j_1$ and $\theta^j_2$ are functions of $r_{\rm sh}$,
$A_j$ also depends on $r_{\rm sh}$. We calculated $A_j$ for the radius
$r_{\rm sh}$ at 2 arcsec intervals.

The actual ejecta shell has a finite width in radius $r_{\rm sh}$. \citet{dec01} 
and \citet{warren05} reported $r_{\rm sh}=190\arcsec-220\arcsec$ for Si, and 
$r_{\rm sh}=180\arcsec-200\arcsec$ for Fe.  We therefore find an average of 
$A_j$ over the given range of $r_{\rm sh}$ as 
\begin{equation}
\label{eq:sme_ave}
\bar{A}_j=\frac{\displaystyle\sum^{r_{\rm sh}^{\rm max}}_{r_{\rm
      sh}=r_{\rm sh}^{\rm min}}\,A_j\,(r_{\rm
    sh})}{\displaystyle\sum^{r_{\rm sh}^{\rm max}}_{r_{\rm sh}=r_{\rm
      sh}^{\rm min}}\,1}.
\end{equation}
In Table \ref{tab:proj_mean}, we summarize the calculated $\bar{A}_j$.

\section{Limited spatial resolution of XRT}

We estimate the smearing effect caused by the limited spatial
resolution of the XRT. We use the {\tt xissim} software, which
simulates interactions of X-ray photons incident upon the XRT/XIS
system using ray-tracing and Monte Carlo techniques
\citep{ishisaki07}.

First, we prepare the input image files for the simulation.  Because
the distributions of K shell emission for IMEs and Fe are slightly
different (e.g., Hwang \& Gotthelf 1997), we prepare two narrow-band
{\it Chandra} images: the Si and S ($1.8-1.9$ plus $2.4-2.5$ keV)
band, and the Fe-K ($6-7$ keV) band. Next, we divide each {\it
  Chandra} image into the SKY regions introduced in Figure
\ref{fig:regions}, and we take this divided image as an input to the
    {\tt xissim} simulator. In the simulation, we fix the input photon
    energy to be monochromatic: 1.8 keV and 6.5 keV for Si+S and Fe,
    respectively. Finally, in the simulated images, we accumulate the
    number of photons in each detector region $1-4$. We repeat this
    procedure for both energy bands (Si+S and Fe) and for all ten
    different SKY regions (SKY 0$-$9).

Results of the simulations are summarized in Table
\ref{tab:sim_results}. We introduce the smearing factor $F_{ij},$
which represents the fraction of the number of photons that originate
in a certain SKY region $j$ ($j=0-9$) relative to to the number of
photons detected in a certain detector region $i$ ($i=1-4$).  For
example, in the Fe K band, $F_{1\,3}=0.194$ means that 19.4\% of
photons detected in region 1 originate from SKY 3.

\section{Correction Factors}

We derive the correction factor $C_i$ by multiplying the projection
factor $\bar{A}_j$ (Table \ref{tab:proj_mean}) by the smearing factor
$F_{ij}$ (Table \ref{tab:sim_results}) as
\begin{equation}
\label{eq:ana_corr}
C_i=\sum_{j=0}^9\,F_{ij}\bar{A}_j~(i=1,2,3,4).
\end{equation}
The derived values are summarized in Table \ref{tab:cf}.
 
\acknowledgments
We gratefully acknowledge all members of the Suzaku hardware and software 
teams and the Science Working Group. A.H. is Research Fellow of Japan 
Society for the Promotion of Science (JSPS). S.K. is supported by JSPS 
Postdoctoral Fellowships for Research Abroad. J.P.H. was supported by NASA 
grant NNG05GP87G.

\clearpage

\begin{deluxetable}{cccccccc}
  \tabletypesize{\scriptsize}
  \tablecaption{Best-fit Gaussian parameters \label{tab:single}} 
  \tablewidth{0pt} 
\startdata 
\tableline\tableline
& \multicolumn{3}{c}{Region 1} & &  \multicolumn{3}{c}{Region 2} \\
    \cline{2-4}\cline{6-8}
Line & Centroid & Width (1$\sigma$) & Flux\tablenotemark{a} && Centroid & Width (1$\sigma$) & Flux\tablenotemark{a}  \\
     & (keV)    & (eV)              &                            && (keV)    & (eV)              &          \\
\tableline
Si He$\alpha$  & 1.8661$^{+0.0002}_{-0.0001}$ & 31.0$\pm0.2$          & 115.7$\pm0.3$              && 1.86603$^{+0.00014}_{-0.00008}$ & 27.0$^{+0.1}_{-0.2}$ & 140.3$^{+0.293}_{-0.301}$ \\
Si Ly$\alpha$  & = He$\beta-0.178$            & = He$\beta$           & 3.5$\pm0.1$                && = He$\beta-0.178$               & = He$\beta$          & 4.94$^{+0.11}_{0.09}$    \\
Si He$\beta$   & 2.1972$^{+0.0012}_{-0.0008}$ & 36$\pm1$              & 10.4$\pm0.1$               && 2.1962$^{+0.0005}_{-0.0008}$    & 30.5$^{+0.6}_{-1}$   & 13.26$^{+0.11}_{0.09}$   \\
Si He$\gamma$  &  = He$\beta+0.111$           & = He$\beta$           & (He$\gamma$/He$\beta$=0.3) && = He$\beta+0.111$               & = He$\beta$          & (He$\gamma$/He$\beta$=0.3)  \\
Si Ly$\beta$   &  = He$\beta+0.193$           & = He$\beta$           & (Ly$\beta$/Ly$\alpha$=0.1) && = He$\beta+0.193$               & = He$\beta$          & (Ly$\beta$/Ly$\alpha$=0.1) \\
\tableline
S  He$\alpha$  & 2.4553$^{+0.0005}_{-0.0004}$  & 40.3$^{+0.5}_{-0.6}$ & 27.4$\pm0.2$               && 2.4549$\pm{0.0003}$ & 36.1$^{+0.3}_{-0.6}$ & 36.3$\pm0.2$  \\
S  Ly$\alpha$  & = He$\beta$-0.263            & = He$\beta$           & 0.18$\pm0.06$              && = He$\beta-0.263$   & = He$\beta$          & 0.395$^{+0.056}_{0.046}$    \\
S  He$\beta$   & 2.899$\pm0.003$               & 59$^{+5}_{-3}$       & 2.05$\pm0.07$              && 2.887$\pm{0.002}$   & 44$^{+2}_{-3}$       & 2.58$\pm{0.05}$     \\
S  He$\gamma$  &  = He$\beta+0.149$            & = He$\beta$          & (He$\gamma$/He$\beta$=0.3) && = He$\beta+0.149$   & = He$\beta$          & (He$\gamma$/He$\beta$=0.3) \\
S  Ly$\beta$   &  = He$\beta+0.222$            & = He$\beta$          & (Ly$\beta$/Ly$\alpha$=0.1) && = He$\beta+0.222$   & = He$\beta$          & (Ly$\beta$/Ly$\alpha$=0.1) \\
\tableline
Ar He$\alpha$  & 3.130$\pm0.002$     & 49$^{+3}_{-2}$ & 2.35$^{+0.07}_{-0.06}$     && 3.129$\pm{0.001}$  & 47$^{+2}_{-1}$ & 3.44$^{+0.06}_{-0.05}$   \\
Ar Ly$\alpha$  &  = He$\alpha+0.209$ & = He$\alpha$   & b                          && = He$\alpha+0.209$ & = He$\alpha$   & b  \\
Ar He$\beta$   &  = He$\alpha+0.561$ & = He$\alpha$   & c                          && = He$\alpha+0.561$ & = He$\alpha$   & c  \\
Ar He$\gamma$  &  = He$\alpha+0.751$ & = He$\alpha$   & (He$\gamma$/He$\beta$=0.3) && = He$\alpha+0.751$ & = He$\alpha$   & (He$\gamma$/He$\beta$=0.3)  \\
Ar Ly$\beta$   &  = He$\alpha+0.812$ & = He$\alpha$   & (Ly$\beta$/Ly$\alpha$=0.1) && = He$\alpha+0.812$ & = He$\alpha$   & (Ly$\beta$/Ly$\alpha$=0.1)  \\
\tableline
Ca He$\alpha$  & 3.850$\pm0.006$     & 82$^{+5}_{-6}$ & 0.78$\pm0.03$              && 3.857$^{+0.004}_{-0.003}$ & 74$\pm{4}$   & 1.13$\pm{0.03}$   \\
Ca Ly$\alpha$  &  = He$\alpha+0.220$ & = He$\alpha$   & b                          && = He$\alpha+0.220$        & = He$\alpha$ & b  \\
Ca He$\beta$   &  = He$\alpha+0.654$ & = He$\alpha$   & c                          && = He$\alpha+0.654$        & = He$\alpha$ & c  \\
Ca He$\gamma$  &  = He$\alpha+0.938$ & = He$\alpha$   & (He$\gamma$/He$\beta$=0.3) && = He$\alpha+0.938$        & = He$\alpha$ & (He$\gamma$/He$\beta$=0.3)  \\
Ca Ly$\beta$   &  = He$\alpha+0.979$ & = He$\alpha$   & (Ly$\beta$/Ly$\alpha$=0.1) && = He$\alpha+0.979$        & = He$\alpha$ & (Ly$\beta$/Ly$\alpha$=0.1)  \\
\tableline
Cr K$\alpha$   & 5.46$\pm0.04$   & = Fe K$\alpha$ & 0.05$\pm0.02$ && 5.51$^{+0.04}_{-0.05}$    & = Fe K$\alpha$ & 0.05$\pm{0.02}$ \\
Fe K$\alpha$   & 6.445$\pm0.004$ & 86$\pm4$       & 1.00$\pm0.03$ && 6.444$^{+0.002}_{-0.003}$ & 71$\pm{3}$    & 1.37$\pm{0.03}$ \\
Fe K$\beta$    & 7.14$\pm0.05$   & = Fe K$\alpha$ & 0.05$\pm0.02$ && 7.11$\pm{0.02}$           & = Fe K$\alpha$ & 0.08$^{+0.02}_{-0.01}$ \\
\tableline\\ 
& \multicolumn{3}{c}{Region 3} & &  \multicolumn{3}{c}{Region 4} \\
    \cline{2-4}\cline{6-8}
Line & Centroid & Width (1$\sigma$) & Flux\tablenotemark{a} && Centroid & Width (1$\sigma$) & Flux\tablenotemark{a}  \\
     & (keV)    & (eV)              &                            && (keV)    & (eV)              &          \\
\tableline
Si He$\alpha$  & 1.86504$^{+0.00006}_{-0.00011}$ & 22.6$^{+0.2}_{-0.1}$ & 182.4$^{+0.4}_{-0.2}$      &&  1.86481$^{+0.00015}_{-0.00007}$ & 20.5$^{+0.1}_{-0.2}$ & 181.2$^{+0.3}_{-0.4}$\\
Si Ly$\alpha$  & = He$\beta-0.178$               & = He$\beta$             & 7.28$^{+0.09}_{-0.13}$     &&  = He$\beta-0.178$               & = He$\beta$          & 7.5$\pm0.1$ \\
Si He$\beta$   & 2.1945$^{+0.0003}_{-0.0007}$    & 23.6$^{+0.6}_{-0.7}$    & 17.69$^{+0.14}_{-0.07}$    &&  2.1935$^{+0.0005}_{-0.0006}$    & 21.1$^{+0.5}_{-1.2}$ & 18.0$\pm0.2$ \\
Si He$\gamma$  &  = He$\beta+0.111$              & = He$\beta$             & (He$\gamma$/He$\beta$=0.3) && = He$\beta+0.111$                & = He$\beta$          & (He$\gamma$/He$\beta$=0.3)  \\
Si Ly$\beta$   &  = He$\beta+0.193$              & = He$\beta$             & (Ly$\beta$/Ly$\alpha$=0.1) && = He$\beta+0.193$                & = He$\beta$          & (Ly$\beta$/Ly$\alpha$=0.1) \\
\tableline
S  He$\alpha$  & 2.4539$\pm{0.0002}$       & 29.9$^{+0.3}_{-0.4}$ & 49.6$^{+0.1}_{-0.2}$       &&  2.4535$^{0.0003}_{-0.0002}$ & 28.7$^{+0.5}_{-0.4}$ & 50.7$^{+0.3}_{-0.2}$ \\
S  Ly$\alpha$  &  =He$\beta-0.263$         & = He$\beta$          & 0.75$^{+0.06}_{-0.05}$      &&  = He$\beta-0.263$           & = He$\beta$          & 0.76$^{+0.09}_{-0.06}$ \\
S  He$\beta$   & 2.888$^{+0.002}_{-0.001}$ & 34$^{+1}_{-2}$       & 3.54$^{+0.06}_{-0.04}$      &&  2.89$^{+0.02}_{-0.01}$      & 35$^{+0.2}_{-0.3}$   & 3.74$^{+0.12}_{-0.06}$ \\
S  He$\gamma$  &  = He$\beta+0.149$        & = He$\beta$          & (He$\gamma$/He$\beta$=0.3) && = He$\beta+0.149$            & = He$\beta$          & (He$\gamma$/He$\beta$=0.3) \\
S  Ly$\beta$   &  = He$\beta+0.222$        & = He$\beta$          & (Ly$\beta$/Ly$\alpha$=0.1) && = He$\beta+0.222$            & = He$\beta$          & (Ly$\beta$/Ly$\alpha$=0.1) \\
\tableline
Ar He$\alpha$  & 3.1273$^{+0.0010}_{-0.0009}$ & 41$^{+1}_{-1}$ & 4.84$^{+0.05}_{-0.06}$     && 3.127$\pm0.01$     & 38$\pm3$     & 5.11$^{+0.08}_{-0.07}$\\
Ar Ly$\alpha$  &  = He$\alpha+0.209$          & = He$\alpha$   & b                          && = He$\alpha+0.209$ & = He$\alpha$ & b  \\
Ar He$\beta$   &  = He$\alpha+0.561$          & = He$\alpha$   & c                          && = He$\alpha+0.561$ & = He$\alpha$ & c  \\
Ar He$\gamma$  &  = He$\alpha+0.751$          & = He$\alpha$   & (He$\gamma$/He$\beta$=0.3) && = He$\alpha+0.751$ & = He$\alpha$ & (He$\gamma$/He$\beta$=0.3)  \\
Ar Ly$\beta$   &  = He$\alpha+0.812$          & = He$\alpha$   & (Ly$\beta$/Ly$\alpha$=0.1) && = He$\alpha+0.812$ & = He$\alpha$ & (Ly$\beta$/Ly$\alpha$=0.1)  \\
\tableline
Ca He$\alpha$  & 3.8612$^{+0.003}_{-0.002}$ & 62$^{+2}_{-3}$ & 1.57$^{+0.03}_{-0.04}$     && 3.868$\pm0.03$     & 49$^{+3}_{-4}$ & 1.52$^{+0.04}_{-0.05}$\\
Ca Ly$\alpha$  &  = He$\alpha+0.220$        & = He$\alpha$   & b                          && = He$\alpha+0.220$ & = He$\alpha$   & b  \\
Ca He$\beta$   &  = He$\alpha+0.654$        & = He$\alpha$   & c                          && = He$\alpha+0.654$ & = He$\alpha$   & c  \\
Ca He$\gamma$  &  = He$\alpha+0.938$        & = He$\alpha$   & (He$\gamma$/He$\beta$=0.3) && = He$\alpha+0.938$ & = He$\alpha$   & (He$\gamma$/He$\beta$=0.3)  \\
Ca Ly$\beta$   &  = He$\alpha+0.979$        & = He$\alpha$   & (Ly$\beta$/Ly$\alpha$=0.1) && = He$\alpha+0.979$ & = He$\alpha$   & (Ly$\beta$/Ly$\alpha$=0.1)  \\
\tableline
Cr K$\alpha$  & 5.48$\pm{0.03}$    & = Fe K$\alpha$ & 0.06$\pm{0.02}$        &&  5.44$^{+0.05}_{-0.04}$    & = Fe K$\alpha$  & 0.05$\pm0.03$\\
Fe K$\alpha$  & 6.448$\pm{+0.002}$ & 58$\pm{3}$     & 1.47$\pm{0.03}$        &&  6.450$^{+0.003}_{-0.004}$ & 53$^{+6}_{-5}$ & 1.09$^{+0.04}_{-0.03}$\\
Fe K$\beta$   & 7.12$\pm{0.03}$    & = Fe K$\alpha$ & 0.05$^{+0.01}_{-0.02}$ &&  7.13$\pm0.09$             & = Fe K$\alpha$  & 0.03$\pm0.02$
\enddata
\tablecomments{Errors indicate the 90\% confidence limits.}
\tablenotetext{a}{Units in $\times10^{-5}$ photons  cm$^{-2}$ s$^{-1}$ arcmin$^{-2}$.}
\tablenotetext{b}{Ly$\alpha$/He$\alpha$ is linked to that of S.}
\tablenotetext{c}{He$\beta$/He$\alpha$ is linked to that of S.}
\end{deluxetable}

\begin{table}
    \begin{center}
    \caption{Correction factors $C_i$ defined in Equation \ref{eq:cf}} 
    \label{tab:cf}
      \begin{tabular}{ccc}
        \hline\hline
        & IMEs  & Fe  \\
        \hline
        Region 1 & 0.748 & 0.712 \\
        Region 2 & 0.578 & 0.530 \\
        Region 3 & 0.315 & 0.262 \\
        Region 4 & 0.170 & 0.120 \\
        \hline
      \end{tabular}
    \end{center}
\end{table}

\begin{table}[!htbp]
  \begin{center}
  \caption{Best-fit parameters of the double Gaussian model}
  \label{tab:double1234}
    \begin{tabular}{ccccccc}
      \hline\hline
      &  Width                     & $E_{\rm red}$ & $E_{\rm blue}$  &  $2\delta E_1$ & $v_{\perp 1}$ & $v_{\rm exp}$ \\
      & (eV)                       & (keV)                           & (keV)                           &  (eV)       & (km s$^{-1}$) & (km s$^{-1}$)\\ 
      \hline
      Si He$\alpha$ & 20.16$^{+0.11}_{-0.09}$ & 1.8883$^{+0.0002}_{-0.0001}$ & $1.8443\pm0.0002$            &  44.0$^{+0.3}_{-0.2}$ & 3540$\pm20$         & 4730$^{+30}_{-20}$   \\  
      Si He$\beta$  & 17.8$^{+0.4}_{-0.7}$    & 2.1685$^{+0.0012}_{-0.0007}$ & $2.2194^{+0.0006}_{-0.0014}$ &  51$^{+1}_{-2}$       & 3480$^{+90}_{-100}$ & 4700$\pm100$ \\ 
      S  He$\alpha$ & 26.6$^{+0.2}_{-0.3}$    & 2.4270$^{+0.0005}_{-0.0004}$ & $2.4841^{+0.0003}_{-0.0005}$ &  57.1$\pm0.6$         & 3490$\pm40$         & 4660$\pm50$ \\
      Ar He$\alpha$ & 34.6$^{+1.4}_{-0.7}$    & 3.089$\pm0.002$              & $3.164\pm+0.002$             &  74$\pm3$             & 3600$\pm100$        & 4800$\pm200$ \\
      \hline
      Fe K$\alpha$  & 55$\pm3$                & 6.507$^{+0.005}_{-0.007}$    & 6.383$^{+0.007}_{-0.005}$    &  124$\pm8$            & 2900$\pm200$        & 4000$\pm300$ \\
      \hline
    \end{tabular}
\tablecomments{Errors indicate the 90\% confidence limits.}
\tablenotetext{a}{Best fit  values of region 1.}
  \end{center}
\end{table}

\begin{table}[!htbp]
  \caption{Calculated $\bar{A}_j$ described in Equation \ref{eq:sme_ave}}
\label{tab:proj_mean}
  \begin{center}
\begin{footnotesize}
    \begin{tabular}{ccccccccccc}
      \hline\hline
  & SKY0 & SKY1 & SKY2 & SKY3 & SKY4 & SKY5 & SKY6 & SKY7 & SKY8 & SKY9\\
\hline 
Si Shell$^{\ast}$ & 0.455 & 0.995 & 0.973 & 0.927 & 0.854 & 0.744 & 0.573 & 0.138 & 0 & 0 \\
Fe Shell$^{\dagger}$ & 0.447 & 0.994 & 0.968 & 0.914 & 0.827 & 0.693 & 0.450 & 0 & 0 & 0 \\
\hline
\multicolumn{11}{l}{\footnotesize$^{\ast}~r_{\rm sh}=190\arcsec-220\arcsec$}\\
\multicolumn{11}{l}{\footnotesize$^{\dagger}~r_{\rm sh}=180\arcsec-200\arcsec$}
    \end{tabular}
\end{footnotesize}
  \end{center}
\end{table}

\begin{table}[!htbp]
    \begin{center}
    \caption{Results of the simulation: fraction of the photons $F_{ij}$ detected in REG $i$ to
      those originated from SKY $j$}
    \label{tab:sim_results}
\begin{footnotesize}
      \begin{tabular}{ccccccccccc}
        \hline\hline
        \multicolumn{11}{c}{Si and S}\\
        \hline
        & SKY 0 & SKY 1 & SKY 2 & SKY 3 & SKY 4 & SKY 5 & SKY 6 & SKY 7 & SKY 8 & SKY 9\\
        \hline
        Region 1 & 0.100 & 0.068 & 0.190 & 0.207 & 0.152 & 0.098 & 0.081 & 0.071 & 0.030 & 0.003 \\
        Region 2 & 0.047 & 0.010 & 0.034 & 0.087 & 0.171 & 0.203 & 0.195 & 0.175 & 0.071 & 0.006 \\
        Region 3 & 0.033 & 0.002 & 0.007 & 0.016 & 0.038 & 0.089 & 0.220 & 0.383 & 0.196 & 0.017 \\
        Region 4 & 0.031 & 0.001 & 0.003 & 0.007 & 0.017 & 0.037 & 0.104 & 0.325 & 0.393 & 0.082 \\
        \hline\hline
        \multicolumn{11}{c}{Fe}\\
        \hline
        & SKY 0 & SKY 1 & SKY 2 & SKY 3 & SKY 4 & SKY 5 & SKY 6 & SKY 7 & SKY 8 & SKY 9\\
        \hline
        Region 1 & 0.105 & 0.064 & 0.178 & 0.194 & 0.154 & 0.113 & 0.099 & 0.062 & 0.024 & 0.006 \\
        Region 2 & 0.050 & 0.009 & 0.032 & 0.079 & 0.164 & 0.223 & 0.235 & 0.143 & 0.054 & 0.013 \\
        Region 3 & 0.037 & 0.002 & 0.006 & 0.016 & 0.039 & 0.100 & 0.271 & 0.335 & 0.160 & 0.036 \\
        Region 4 & 0.036 & 0.001 & 0.003 & 0.006 & 0.016 & 0.040 & 0.120 & 0.261 & 0.341 & 0.176 \\
        \hline
      \end{tabular}
\end{footnotesize}
    \end{center}
\end{table}

\end{document}